\documentstyle[12pt,aaspp4]{article}

\lefthead{Bekki \& Couch}
\righthead{Starbursts by intracluster hot gas}

\received{2003 April 22}
\begin{document}
\title{Starbursts from strong  compression of galactic molecular  clouds
due to the high pressure of the intracluster medium}

\author{Kenji Bekki \& Warrick J. Couch} 
\affil{
School of Physics, University of New South Wales, Sydney 2052, Australia}

\begin{abstract}
We demonstrate that the
high pressure of the hot intracluster medium (ICM) can trigger the collapse
of molecular clouds in a  spiral galaxy, leading to a burst of star formation
in the clouds. 
Our hydrodynamical simulations show that the 
high gaseous (ram pressure and static thermal) pressure of the ICM strongly
compresses a self-gravitating  gas cloud within a short time scale ($\sim$
$10^{7}$ yr), dramatically 
increasing the central gas density,
and consequently causing efficient star formation within the cloud.
The stars developed in the cloud  form a compact, gravitationally bound, star cluster.
The star formation efficiency within such a cloud 
is found to depend on the temperature and the density of the ICM
and the relative velocity of the galaxy with respect to it.
Based on these results, 
we discuss the origin of starburst/poststarburst populations observed in distant
clusters,  the enhancement of star formation for  galaxies in merging clusters, 
and the isolated compact HII regions recently discovered in the Virgo cluster. 
\end{abstract}


\section{Introduction}

Recent spectrophotometric and morphological studies of cluster galaxies
by large ground--based telescopes and the $Hubble$ $Space$ $Telescope$ ($HST$)
have revealed that the star formation histories 
are very different between different morphological and luminosity classes
and can be closely associated with
galactic morphological evolution driven by cluster-related
physical processes (e.g., Lavery \& Henry 1994; Couch et al. 1994, 1998;
Abraham et al. 1996; Balogh et al 1999;  Dressler et al. 1999).
For example, Couch et al. (2001)  revealed that star formation
of disk galaxies in distant clusters are globally and uniformly suppressed.
Poggianti et al. (1999) found  that
galaxies with ``e(b)'' spectral type,
which are the most likely to be  starbursting galaxies,
generally belong to the low-luminosity dwarf populations in clusters.
It has been a longstanding and remarkable problem 
what mechanisms are responsible for the observed enhancement/decline of
galactic star formation rates in clusters.

The dynamical and hydrodynamical effects of the hot, high-temperature
ICM have long been suggested to play decisive roles, not only
in transforming galactic morphological properties
but also in  controlling  star formation rates in galaxies
within clusters (e.g., Gunn \& Gott 1972;
Gavazzi \& Jafffe 1985; Bothun\& Dressler 1986;
Evrard 1991). 
Fujita \& Nagashima (1999) demonstrated that 
star formation rates and photometric  properties of galaxies in clusters 
can be changed as a result of the ram pressure effects on galactic  molecular
clouds.
Using three dimensional SPH/N-body simulations,
Abadi et al. (1999) found  that although the ram pressure of the ICM
can efficiently strip interstellar HI gas from spirals,
such ram pressure stripping alone can not abruptly truncate star formation in
spirals. However, 
these previous phenomenological/numerical models
have not investigated the effects of the ICM on 
{\it individual molecular clouds with internal structures,
kinematics, and chemical abundance}, even though star formation
is ongoing within individual molecular clouds.  
Therefore, it is still unclear 
how the star formation rate and  efficiency in individual galactic molecular
clouds is influenced by the hot ICM 
(and thus how the galactic star formation can be controlled by the ICM).

The purpose of this {\it Letter} is to first demonstrate that the 
high gaseous  pressure of the ICM significantly changes 
the internal structure of a self-gravitating molecular gas cloud in a cluster
disk galaxy and consequently trigger a burst of star formation
within the cloud. In particular, 
we investigate (1)\,the star formation caused 
within gas clouds in a disk galaxy due to the strong compression of the gas by
the ICM, 
and (2)\,its dependence on the density and the temperature of the ICM 
and the velocity of the galaxy relative to the ICM.
We show  that both ram pressure and static pressure from the ICM 
on the galactic gas clouds are important in compressing the clouds and thus in
triggering starbursts within the clouds.  
We emphasize that although the ICM can be   
responsible for the stripping of {\it HI gas within disks} and {\it diffuse halo
gas}  in  cluster galaxies 
and thus for the truncation of star formation  (e.g., Abadi et al. 1999; Bekki
et al. 2002), it  could also be closely associated with  
starburst activity in cluster/group environments. Finally, this proposed new
mechanism for triggering starbursts is discussed in a variety of 
different contexts of cluster galaxy evolution.

\placefigure{fig-1}
\placefigure{fig-2}

\section{Our Model}
By using TREESPH code with star formation methods (Bekki 1997),
we numerically investigate the hydrodynamical effects of the ICM 
on a self-gravitating molecular gas cloud
orbiting within  a spiral galaxy under the gravitational influence of the galaxy
and its surrounding dark matter halo.
The  cloud is represented by 20,000 SPH particles and
the initial cloud mass ($M_{\rm cl}$) and  size ($r_{\rm cl}$)
are set to be $10^{6}\,M_{\odot}$ and 97\,pc, respectively,
which are consistent
with the mass-size relation observed by Larson (1981).
The  cloud is assumed to have an isothermal radial density profile
with $\rho (r) \propto 1/(r+a)^2$, where $a$ is the core radius of the cloud
and set to be $0.2r_{\rm cl}$.
An isothermal equation of state with a sound speed of $c_{\rm s}$
is used for the gas, and $c_{\rm s}$ is set to be 4\,km\,s$^{-1}$
(consistent with the prediction from the virial theorem)
for models with $M_{\rm cl} = 10^{6}\,M_{\odot}$.

Each SPH particle is subject to the gravitational forces from the {fixed} spiral
potential
that is assumed to have three components: a dark
matter halo, a disk, and a bulge. We assume a logarithmic dark matter halo
potential
(${\Phi}_{\rm halo}=v_{\rm halo}^2 \ln (r^2+d^2)$),
a Miyamoto-Nagai (1975) disk
(${\Phi}_{\rm disk}=-GM_{\rm disk}/\sqrt{R^2 +{(a+\sqrt{z^2+b^2})}^2}$),
and a spherical Hernquist (1990) bulge 
(${\Phi}_{\rm bulge}=-GM_{\rm bulge}/(r+c)$),
where $r$ is the distance from the center of the spiral, 
$d$ = 12\,kpc, $v_{\rm halo}$ = 131.5\,km ${\rm s}^{-1}$,
$M_{\rm disk}$ =  $10^{11}$ $M_{\odot}$,
$a$ = 6.5\,kpc, $b$ = 0.26\,kpc, $M_{\rm bulge}$ =  3.4 $\times$ $10^{10}$ 
$M_{\odot}$, and $c$ = 0.7\,kpc. This set of parameters is reasonable and 
realistic for the Galaxy. The cloud
is initially within the disk plane (corresponding to the $x$-$y$ plane)
and located at ($x$, $y$, $z$) =  ($R_{g}$, 0, 0).
The initial velocity of the cloud is set to be ($v_{\rm x}$, $v_{\rm y}$, $v_{\rm z}$)
= (0, $v_{\rm c}$, 0), where $v_{\rm c}$ is the circular velocity at the position.
We show the results for $R_{g} = 8.5$\,kpc (corresponding to the solar neighborhood)
in this paper,
because we find that 
the star formation within clouds does not depend strongly on  $R_{g}$.

Each SPH particle is also subject to the hydrodynamical force of the ICM whose
strength depends on the parameters of the ICM. The ICM is represented by 
8,824 SPH particles and 
and an isothermal equation of state with temperature, $T$ (or sound velocity
of $c_{\rm h}$),  and density, $\rho$. 
The cloud -- which is in a spiral that orbits the cluster core with a 
velocity, $V_{\rm rel}$, relative to the ICM --  
is subject to both ram pressure 
($P_{\rm ram}$ $\propto$  $\rho {V_{\rm rel}}^{2}$) and thermal static
pressure ($P_{\rm s}$ $\propto$ $\rho {c_{\rm h}}^{2}$). 
The ICM particles are uniformly distributed in an annular ring with size 
$R_{\rm g}$, width $6R_{\rm cl}$, and thickness $6R_{\rm cl}$
(in the $z$ direction) on the disk. 
We give all ICM particles an initial velocity of (0, $V_{\rm rel}$, 0), 
so that we can represent the hydrodynamical effects that the ICM has
on the spiral when it passes through the cluster core. 
Periodic boundary conditions are adopted only for these ICM particles.
We here stress that for the adopted total number of SPH particles ($\sim$30000),
the simulations can not fully resolve the contact surface between
the cold GMC and the hot ICM (accordingly thermal conduction between
these gas and the cloud evaporation can not be fully investigated).
The star formation rate in our simulations can be therefore overestimated
(as discussed later in this paper).

A gas particle in a cloud is converted into a collisionless stellar particle 
if (1)\,the local dynamical time scale [corresponding to  
${(4 \pi G\rho_{i})}^{-0.5}$, where $G$ and $\rho_{i}$ are the gravitational
constant and the density of the gas particle, respectively]
is shorter than the sound crossing time (corresponding to $h_{i}/c_{\rm s}$, 
where $h_{i}$ is the smoothing length of the gas), 
and (2)\,the gaseous flow is converging.
This method thus mimics star formation due to Jeans instability in gas clouds. 
In the model without  ICM effects, star formation does not occur at all in this
star formation method.
By changing the values of the three parameters 
(i.e., $T$, $\rho$, and $V_{\rm rel}$),
we investigate the effects of the ICM on the cloud. 
We show the results for the models 
with  $0.01 T_{0}$ $\le$ $T$ $\le$ $T_{0}$,
with  $0.1 {\rho}_{0}$ $\le$ $\rho$ $\le$ ${\rho}_{0}$,
and  $0.1 {V}_{0}$ $\le$ $V_{\rm rel}$ $\le$ ${V}_{0}$,
where $T_{0}$, ${\rho}_{0}$, and ${V}_{0}$
are 8 keV, 5.6 $\times$ $10^{-27}$ g cm$^{-3}$, and 1000 km s$^{-1}$. 
These values of $T_{0}$ and ${\rho}_{0}$ correspond to
the observed values of the ICM in the central region of the Coma cluster
(Briel et al. 1992).
We mainly   show the result of the ``fiducial model''
with $T$ = $T_{0}$, $\rho$ = $0.1 {\rho}_{0}$, and $V_{\rm rel}$ = 1000  km s$^{-1}$,
because this model shows the typical behavior of ICM-induced star formation
in a gas cloud. The gas cloud can not be suddenly stripped because of the higher initial 
column density during simulations (See also Fujita \& Nagashima 1999).

\section{Results}

Figures 1 and 2 describe how a burst of star formation within a gas cloud
can be triggered by the external gas pressure of the ICM in the fiducial model.
In this model with moderately strong ram pressure,
the high pressure of the ICM can continue to strongly compress 
the cloud without losing a significant amount of gas from the cloud
(i.e., only 3\% of the initial mass can be stripped within 14\,Myr). 
As the strong compression proceeds, 
the internal density/pressure of the cloud can rise significantly.
However, the self-gravitational force of the cloud also becomes stronger
because the cloud becomes progressively more compact during the ICM's compression. 
Therefore, the internal gaseous pressure of the cloud
alone becomes unable to support itself against the combined effect of
the external pressure from the ICM and the stronger self-gravitational force.
As a result of this,
the cloud's collapse initially induced by the high external pressure from the 
ICM can continue in a runaway manner.

Due to the rapid, dissipative collapse, 
the gaseous density of the cloud dramatically rises and 
consequently star formation begins in the central regions of the cloud.
The star formation rate increase significantly from $0.1$ $M_{\odot}$ yr$^{-1}$ 
(for the first 5\,Myr) to 
0.6 $M_{\odot}$ yr$^{-1}$ (8\,Myr after the start of the cloud's collapse).
This corresponds to global galactic star formation rate of 
$30-60$ $M_{\odot}$ yr$^{-1}$ for galaxies with total H$_2$ mass similar
to that of the Galaxy.
About 81\% of the gas is converted into stars within 14\,Myr
to form a stellar system. Because of the ``implosive'' formation of stars from
strongly compressed gas, the developed stellar system is strongly
self-gravitating and compact. This result implies that high external pressure
from the ICM is likely to trigger the formation of bound, compact star
clusters rather than unbound, diffuse field stars.

Parameter dependences are found to be complicated and can be summarized as
follows (see Figure 3): Firstly, for the high ICM temperature ($T$ = $T_{0}$), 
large relative velocity ($V_{\rm rel}$ = $V_0$) models, 
the star formation efficiency (hereafter SFE) within a cloud
is higher for the model with the lower ICM density ($\rho$). 
This is because in the higher $\rho$ model, the ram pressure 
of ICM strips a larger amount of gas from the cloud and thus decreases   
the total amount of gas that can be converted into stars.
Secondly, for the models with lower $\rho$ (= $0.1 {\rho}_{0}$)
and lower $T$ (= $0.1 {T}_{0}$), the SFE is higher for the model with 
the larger $V_{\rm rel}$. This suggests that the moderately strong ram pressure
of the ICM can play a role in enhancing the star formation rate in a gas cloud.
Thirdly, for models in which ram pressure has a negligible effect on the gas
cloud (i.e, high density, $\rho$ =${\rho}_{0}$, and very low ICM
relative velocity, $V_{\rm rel}$ = 0.01$V_0$), the SFE is likely to be higher
for the model with higher $T$. This is because the higher static (thermal,
external) pressure of the ICM in the model with higher $T$
can more strongly compress a cloud and thus produce regions of high gaseous
density within. Fourthly, for models in which ram pressure is very strong
($\rho$ = ${\rho}_{0}$ and $V_{\rm rel}$ = $V_0$),
there is no clear dependence on $T$.

\section{Discussion and conclusions}

By assuming that galactic molecular clouds can be exposed to the hot ICM, 
we have demonstrated that the high pressure from the ICM can trigger 
efficient star formation in gas clouds in cluster spirals. 
However, the diffuse outer halo gas and the disk HI gas of a cluster spiral
could prevent the molecular gas clouds from being directly exposed to 
the hot ICM. 
Using semianalytic models of galaxy formation,
Okamoto \& Nagashima (2003) demonstrated that molecular gas clouds
can be consumed up by star formation  in cluster spirals before the ram pressure effects
become significant for their evolution. 
Furthermore, evaporation of gas clouds by heat conduction (Cowie \& McKee 1977) could
reduce significantly the gas mass that can be converted into stars,
because the time scale of cloud evaporation is an order of $10^6-10^8$ years 
for the adopted parameters of gas clouds in the present study.
For the adopted cloud parameters, the fraction of gas that
can be evaporated before star formation within the gas  can be estimated as 12 \%.
Therefore, the proposed ICM-induced starbursts in molecular gas
clouds may be more likely for galaxies which only have small amounts of
HI and halo gas yet have enough molecular gas. 
In spite of this strong limitation in the applicability of the
proposed mechanism to the starburst phenomenon within clusters,
we suggest that the present results may still have the following three
important implications for cluster galaxy evolution.

The first is that if the ICM triggers a starburst in a disk galaxy
rich in molecular gas, actively star-forming regions can be widely spread
throughout the disk. The expected distribution of starburst regions
within a disk is in striking contrast to the nuclear starbursts
suggested to be triggered by tidal interactions/mergers between galaxies
and global cluster tidal fields (e.g., Byrd \& Valtonen 1990). 
Furthermore, the starburst mechanism proposed here for cluster disk galaxies
differs from that for mergers in that the thin disk component 
remains intact after the starburst in the former whereas it is destroyed
to form a thick disk in the latter (e.g., Bekki 1998).
Therefore, {\it widely spread starburst (or poststarburst) activity 
within the thin disk component} of a cluster disk galaxy is the signature
to look for in identifying this ICM-gas cloud interaction.
Spectroscopic observations that can reveal the detailed distributions
of starburst/poststarburst regions within a disk
(e.g., the GMOS Integral Field Unit on the 8m Gemini telescope) will reveal such
ICM-induced starburst/poststarburst populations in distant clusters.

The second is that a larger fraction of starburst/poststarburst
populations are expected in merging clusters, where a higher ICM temperature
(and pressure) is predicted  (e.g., Roettiger et al. 1997). 
However, it is unclear (both observationally and theoretically)
whether cluster merging triggers starbursts in cluster disk galaxies
and what mechanisms are responsible for the formation of starburst
galaxies in merging clusters. Although some observational evidence has
been found for merging clusters having a larger number of
starbursts/poststarburst candidates both at low (Caldwell  et al. 1993; Caldwell
\& Rose 1997) and intermediate redshifts  (e.g., Owen et al. 1999),
Venturi et al. (2000) found no such clear evidence
in the merging clusters Abell 2125 and 2645.
Bekki (1999) demonstrated that the time-dependent tidal gravitational
field of merging clusters can trigger starbursts in their galaxies
whereas Fujita et al. (1999) concluded that ram pressure stripping
can strongly suppress the star formation of galaxies in merging clusters.
More extensive observational investigations of (1)\,a correlation between the
fractional content of starburst/poststarburst galaxies and the
incidence/non-incidence of cluster merging, and (2)\,the spatial distribution
of starburst/poststarburst galaxies within merging clusters, 
will address the above two unresolved problems in a more quantitative way.

The third is that the origin of isolated intracluster
compact HII regions recently discovered in the Virgo cluster 
(e.g., Yoshida et al. 2002; Arnaboldi et al. 2003; Gerhard et al. 2003)
can be closely associated with  ICM-induced efficient star formation
in gas clouds drifting in the cluster. The ICM-induced efficient star formation
might be highly likely to be seen in the intracluster molecular gas clouds 
stripped from cluster spirals through interactions between them (and between
them and the cluster global tidal fields),
because these intracluster gas clouds are directly exposed to ICM.
We suggest that the total number and spatial distribution of isolated
intracluster HII regions in the Virgo cluster provide valuable information
on the cluster's dynamical history of galaxy interaction/merging that have been
responsible for stripping of molecular clouds within disks.
Our results also imply that the formation of intracluster globular clusters
(West et al. 1995) and planetary nebulae (Arnaboldi et al 1996) could result
from  the past interaction between stripped gas clouds and ICM.

The present results imply that some fraction of molecular clouds in cluster spirals 
could disappear from their disks, not by tidal or ram pressure stripping but by 
rapid consumption due to  ICM-induced star formation.
They also imply that the initial mass function (IMF) of a gas cloud with
ICM-induced star formation could be different from the normal Salpeter-like IMF,
because the ICM-induced star formation can occur preferentially in the central
regions of gas clouds, where more massive stars are likely
to be formed (e.g., Murray et al. 1996). Our future, more sophisticated
simulations which will include chemical evolution,
magnetic fields,  dynamical evolution of hierarchical/fractal  structures within
a cloud, and feedback effects from massive stars and supernovae will
address the total amount of molecular gas consumed by ICM-induced star formation
and its IMF in a more quantitative way.

\placefigure{fig-3}
\placefigure{fig-4}

\acknowledgments
We are  grateful to the anonymous referee for valuable comments,
which contribute to improve the present paper. 
K.B. and W.J.C. acknowledge the financial support of an Australian Research
Council Discovery Program grant throughout the course of this work.

\clearpage


\figcaption{
Distribution of gas (cyan) and new stars formed from gas (magenta)
of the self-gravitating cloud projected onto the $x$-$z$ plane  for  the fiducial model,
at each time indicated in the upper left corner of each panel.
The size is given in units of the cloud size (97 pc) so that
each frame measures 423 pc on a side.
the upper left corner of each panel.  
The upper three short arrows 
and the lower long arrow  
indicate the direction of the ICM flow with respect to the galactic motion
and that of the cloud motion, respectively.
The cloud is initially within the $x$-$z$ plane plane and orbiting
within the disk in a counter-clockwise.
Note that owing to strong external compression by ICM,
the gas cloud rapidly collapses to start star formation
in its central region.
\label{fig-1}}

\figcaption{
Time evolution of the star formation rate in units of $M_{\odot}$ yr$^{-1}$
(upper) and  that of the normalized gas mass (solid) and stellar mass (dotted).
Here gaseous and stellar masses are normalized to initial gas mass. 
\label{fig-2}}

\figcaption{
Dependence of the mass fraction of new stars 
($M_{\rm s}/M_{\rm g}$) within a cloud  on
model parameters: Dependences on ICM density (upper left),
relative velocity between ICM and disk galaxies (upper right),
and ICM temperature (lower left and  right).
$M_{\rm s}$ and $M_{\rm g}$ are the total mass of new stars
formed within  14 Myr and initial gas mass of the clouds, respectively.
In each panel,  four models with different values
of one of the three parameters ($\rho$, $T$, and $V_{\rm rel}$)
and the fixed  ones of the other two
are plotted.
The fixed parameter values of temperature, density, and relative
velocity (represented by T, D, and V) are indicated in the
upper right corner of each panel in units of 
$T_{0}$ (8 keV),
${\rho}_{0}$ (5.6 $\times$ $10^{-27}$ g cm$^{-3}$), 
and $V_{0}$ (1000  km s$^{-1}$), respectively.
For example, the upper left panel shows
the dependence of $M_{\rm s}/M_{\rm g}$ on ICM density
for models with $T$ = 8 keV, $V_{\rm rel}$ = 1000  km s$^{-1}$,
and $\rho$ = 0.1, 0.25, 0.5, 1.0${\rho}_{0}$. 
\label{fig-3}}

\newpage
\plotone{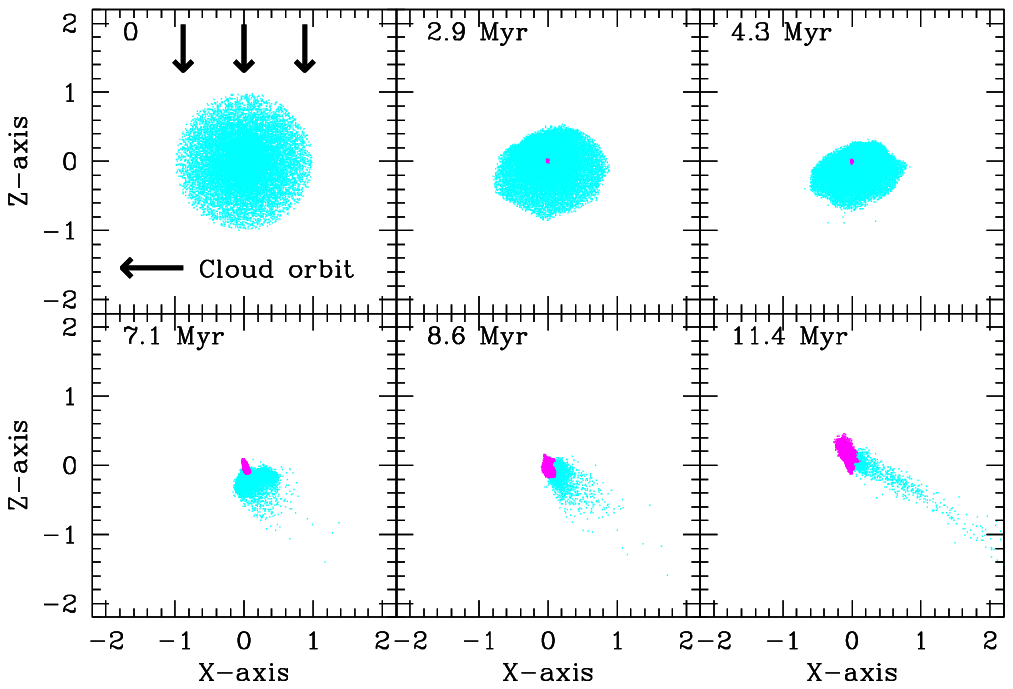}
\newpage
\plotone{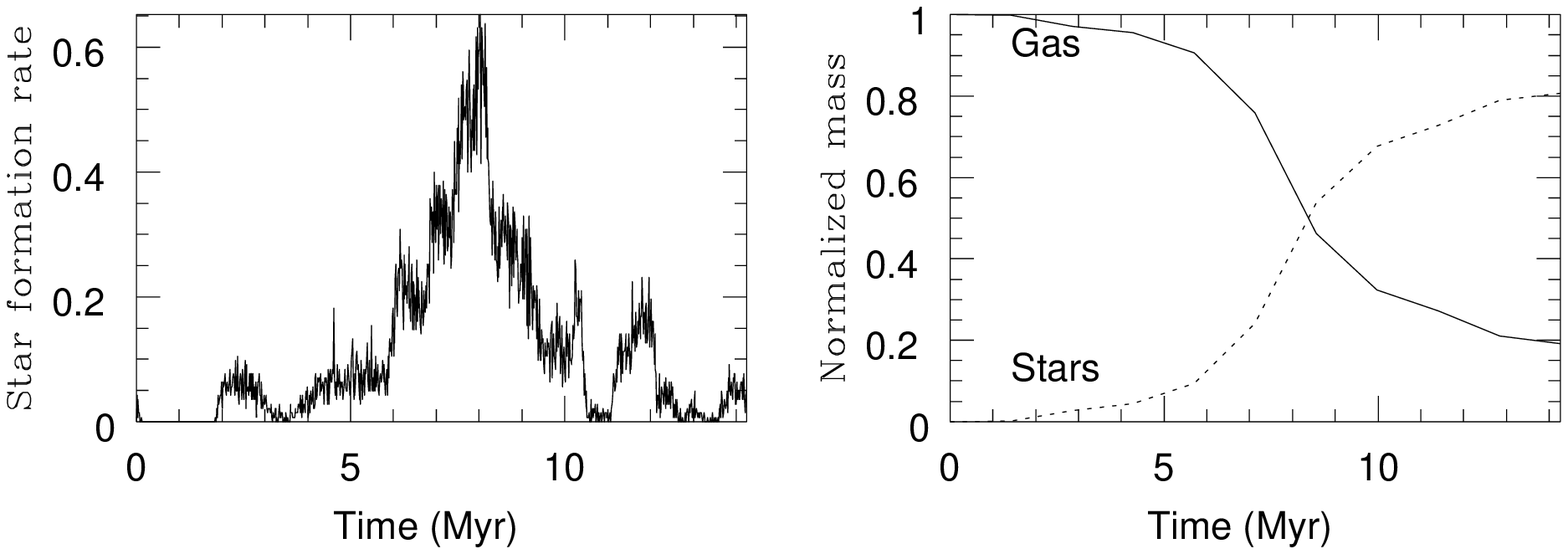}
\newpage
\plotone{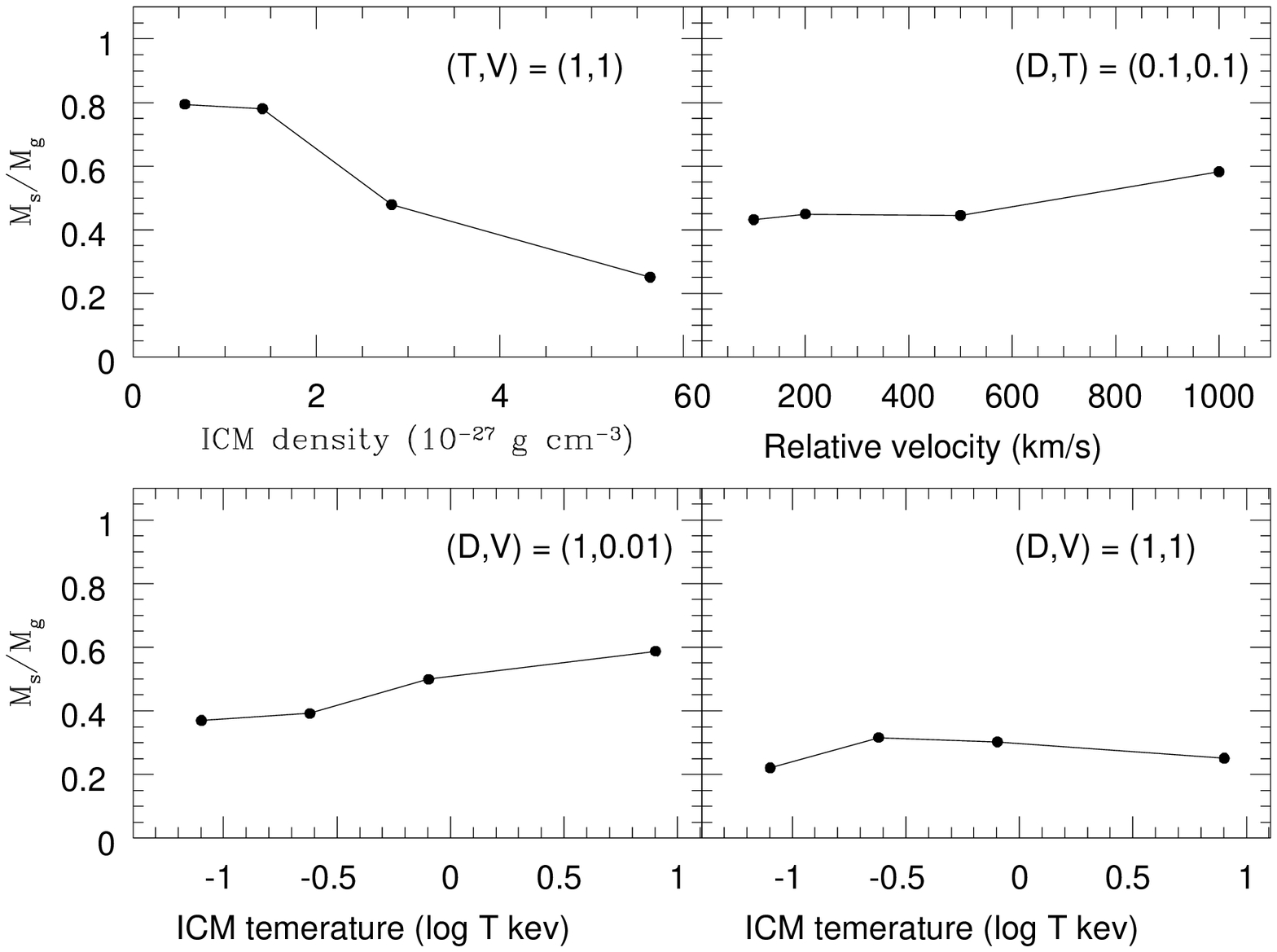}

\end{document}